\documentclass{article}

\usepackage{PRIMEarxiv}

\usepackage[utf8]{inputenc} 
\usepackage[T1]{fontenc}    
\usepackage{hyperref}       
\usepackage{url}            
\usepackage{booktabs}       
\usepackage{amsmath,amssymb,amsfonts} 
\usepackage{nicefrac}       
\usepackage{microtype}      
\usepackage{fancyhdr}       
\usepackage{graphicx}       
\graphicspath{{media/}}     

\usepackage{algorithm}
\usepackage{algorithmic}

\usepackage{graphicx}
\usepackage{wrapfig}

\title{Empowering Future Cybersecurity Leaders: Advancing Students through FINDS Education for Digital Forensic Excellence}

\author{
  \textbf{Yashas Hariprasad$^{1}$, Subhash Gurappa$^{2}$, Sundararaj S. Iyengar$^{2}$,} \\
  \textbf{Jerry F. Miller$^{3}$, Pronab Mohanty$^{4}$, Naveen Kumar Chaudhary$^{5}$} \\
  \\
  $^{1}$California State University, East Bay, CA, USA \\
  \texttt{yashas.hariprasad@csueastbay.edu} \\
  \\
  $^{2}$Florida International University, Miami, Florida, USA \\
  \texttt{\{sg001, iyengar\}@fiu.edu} \\
  \\
  $^{3}$Florida Agriculture and Mechanical University, Tallahassee, Florida, USA \\
  \texttt{jerry.miller@famu.edu} \\
  \\
  $^{4}$Director General of Police (DGP), Police Computer Wing, Indian Police Service, Govt. of India \\
  \\
  $^{5}$National Forensic Sciences University, Gandhinagar, Gujarat, India \\
  \texttt{naveen.chaudhary@nfsu.ac.in}
}

\begin{document}
\maketitle

\begin{abstract}
The Forensics Investigations Network in Digital Sciences (FINDS) Research Center of Excellence (CoE), funded by the U.S. Army Research Laboratory, advances Digital Forensic Engineering Education (DFEE) through an integrated research–education framework for AI-enabled cybersecurity workforce development. FINDS combines high-performance computing (HPC), secure software engineering, adversarial analytics, and experiential learning to address emerging cyber and synthetic media threats.

This paper introduces the Multidependency Capacity Building Skills Graph (MCBSG), a directed acyclic graph–based model that encodes hierarchical and cross-domain dependencies among competencies in AI-driven forensic programming, statistical inference, digital evidence processing, and threat detection. The MCBSG enables structured modeling of skill acquisition pathways and quantitative capacity assessment.

Supervised machine learning methods, including entropy-based Decision Tree Classifiers and regression modeling, are applied to longitudinal multi-cohort datasets capturing mentoring interactions, laboratory performance metrics, curriculum artifacts, and workshop participation. Feature importance analysis and cross-validation identify key predictors of technical proficiency and research readiness.

Three-year statistical evaluation demonstrates significant gains in forensic programming accuracy, adversarial reasoning, and HPC-enabled investigative workflows. Results validate the MCBSG as a scalable, interpretable framework for data-driven, inclusive cybersecurity education aligned with national defense workforce priorities.
\end{abstract}


\section*{Sponsorship and Disclaimer}
This research was sponsored by the Army Research Office and the NSF, and was accomplished under Grant Number W911NF-21-1-0264 and 2018611. The views and conclusions contained in this document are those of the authors and should not be interpreted as representing the official policies, either expressed or implied, of the Army Research Office or the U.S. Government. The U.S. Government is authorized to reproduce and distribute reprints for Government purposes notwithstanding any copyright notation herein.

\section{Introduction}
\label{sec:introduction}
The pervasive growth and widespread use of digital devices in our society, government, and military have raised concerns about the vulnerability to malfeasance, cyberattacks, and illegal penetration of devices, exposing sensitive information to adversaries \cite{b1}. In the aftermath of such incidents, the forensic process becomes crucial. The demand for expertise, tools, and techniques in digital forensics is now more critical than ever, and this demand is expected to skyrocket with the rise of autonomous vehicles, the increased use of mobile devices, drones, and devices connected to the Internet of Things \cite{b2}. Coupled with the rapid growth of computer espionage and cybercrime, the Bureau of Labor Statistics predicts a 32\% growth in digital forensics examiners armed with advanced tools and techniques by 2028 \cite{b3}.

Addressing this need, Florida International University (FIU), in collaboration with three Historically Black Universities: Grambling State University, Jackson State University, Florida A\&M University, and various other global collaborations, established a groundbreaking Digital Forensics Research Center of Excellence. Named, the Forensics Investigations Network in Digital Sciences (FINDS) Research Center of Excellence (COE), this initiative funded by the US Army Research Laboratory (ARL) since 2021, is focusing on advanced research and education in emerging digital forensic areas with the intersection of science education and technology (https://finds.fiu.edu/). The collaboration extends to DoD (Department of Defense) agencies, industry partners, international collaborators, and national laboratories, fostering research, technology, and mentoring. For a broader treatment refer to our annual reports. 

The FINDS Center of Excellence conducts extensive research in digital forensics, focusing on three key theme areas: \textit{AI-powered Analytical Methods/Evidence Processing Techniques, Forensic Fusion Models for Extracting Event Signatures, and Drone Forensics and Ubiquitous Forensic Signatures}. A crucial aspect of this center involves fostering workforce development by integrating a diverse range of mentors dedicated to ensuring a student's research experience is reflective and supportive of diversity \cite{b4}. This method stands out for its inclusion of mentors from both research and government laboratories, offering extensive applications in national security. Students are exposed to a diverse range of digital forensics techniques, with a particular focus on integrating advanced Artificial Intelligence/Machine Learning (AI/ML) methods to markedly elevate progress in applied forensic sciences. 

The Center of Excellence (CoE) has built a strong educational foundation in Digital Forensics, with a specific focus on recognizing inventive directions and tools created by students and faculty from ethnic minority backgrounds, encompassing Black students, Hispanic students, and women. The impact of the results obtained thus far has significantly propelled the careers of these individuals in STEM (Science, Technology, Engineering, and Math) disciplines. Since its inception, FINDS has sponsored and hosted four annual digital forensic workshops, providing training to over 400 students worldwide. Additionally, advanced graduate courses were offered, and various projects were successfully published in reputable journals.

The success of this initiative extends beyond national borders, aiming to make a significant contribution to capacity building in the global engineering community. By establishing a robust pathway to academia through a rigorous cybersecurity and digital forensics graduate education program, FINDS envisions shaping the future generation of cyber warriors. Utilizing its extensive student body in science, engineering, and computing, FINDS can create living laboratories for cutting-edge research, fostering a dynamic exchange between research, practical application, and curriculum enhancement. 

This paper presents a Multi-Dependency Capacity Building Graph Data Structure, which was utilized by FINDS to effectively train emerging cyber warriors and enhance skills in digital forensics and cyber security. This framework underwent thorough evaluation and testing using real-world datasets that encompass over three years of data, employing methods such as Decision Tree Classifiers and Statistical Modelling. The results of the experiments, derived from data analysis and student feedback, show remarkable success in advancing educational capabilities in Digital Forensic Engineering. The preliminary results suggest that FINDS has successfully produced scholars and leaders whose impact reaches far beyond geographical boundaries.

\textit{\textbf{Summary of Contribution:}}
The outcomes of this work are manifold and have significant implications for the field of digital forensics and cybersecurity:
1. Development of a novel multi-dependency capacity-building graph framework for optimizing resource allocation and workforce development.
2. Introduction of advanced AI/ML methodologies for enhancing digital forensic education and practice.
3. Demonstration of the effectiveness of the proposed framework through validation and case studies.
4. Establishment of pathways for underrepresented groups to access high-quality education and career opportunities in digital forensics and cybersecurity.

\textit{\textbf{Organization of the Paper:}}
The structure of the paper unfolds as follows: Section 2 delves into the motivational paradigm driving capacity building. Section 3 outlines the Essential Imperative, focusing on cultivating a robust workforce through development initiatives. In Section 4, we explore the Frontiers of Innovation, elucidating how FINDS' technical advancements elevate AI-powered digital forensics. Section 5 presents the Results and Discussions, providing insights into the impact of FINDS based on real-time data. Section 6 encapsulates the concluding remarks and acknowledgments.

\section{Motivation of the Work}
The rapid proliferation of digital technologies has brought unprecedented opportunities but also significant challenges in the realm of cybersecurity and digital forensics. As cybercrimes escalate globally, there is a pressing need to enhance the capacity of digital forensic professionals to tackle increasingly sophisticated threats. This work is motivated by the critical need to develop innovative tools and frameworks that bridge the gap between theoretical advancements and practical applications, fostering a resilient workforce capable of addressing the demands of the digital age. Furthermore, this research seeks to address the specific challenges faced by underrepresented communities in accessing cutting-edge education and career opportunities in digital forensics and cybersecurity.

\subsection{Motivation Paradigm - Iyengar's Law: Tripling of Cybercrimes Incidents Every 3 Years}
In the interconnected world of the 21st century, the rise of cybercrimes has become an undeniable reality, and Iyengar's Law encapsulates a troubling trend - the tripling of cybercrime incidents every three years in major cities globally \cite{b5}. This phenomenon is acutely exemplified, and the relentless march of technology has become a double-edged sword, fostering both progress and peril.

\subsection{Exploring Facts and Observations: A Closer Look at Understanding and Analyzing Cyber Information}
\begin{figure}
    \centering
    \includegraphics[scale=0.7]{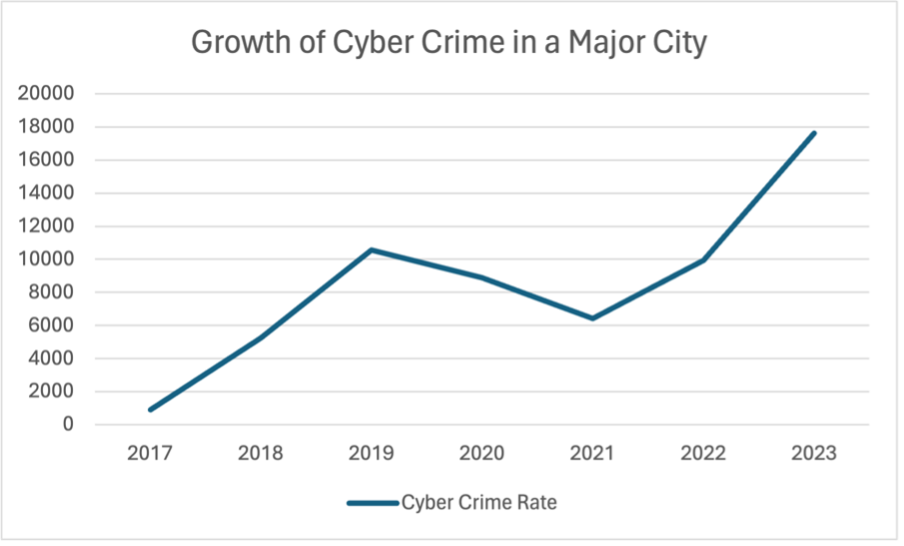}
    \caption{Exponential Growth Rate of Cyber Crime in a Major City, \textit{Adopted from: https://indianexpress.com/article/cities/bangalore/cybercrime-cases-recorded-2022-bengaluru-secures-top-spot-9055838/}}
    \label{fig1}
\end{figure}

In the global cyber landscape, the numbers present a compelling narrative. Data collected from major cities worldwide reveals a consistent upward trend. For instance, in 2021, a major city reported 6,422 cybercrimes, signaling the onset of a growing threat. By 2022, the number of reported cases surged significantly to 9,940. In 2023, the situation escalated further, with the city contending with an alarming 17,623 reported cybercrimes, as illustrated in Figure.\ref{fig1}. 

This stark progression mirrors the predicted trajectory of Iyengar's Law, emphasizing the urgent need for a robust response to this escalating digital threat. The driving force behind this surge lies in the relentless pace of technological development \cite{b6}. While innovations empower societies, they also empower cybercriminals, providing them with increasingly sophisticated tools and methodologies. As technology blurs geographical boundaries, the impact is felt globally, and in cities like Bangalore, where technologically savvy criminals can form a nexus of opportunity and financial need leading to an explosion of the digital criminal underworld.

Critical to understanding and addressing this surge is the role of cyber hotlines and streamlined case registration through police hotlines. These mechanisms not only serve as essential tools for responding to cybercrimes but also act as proactive shields, enabling authorities to anticipate, prevent, and combat emerging threats effectively. Beyond the specific case, global cybersecurity challenges persist. The interconnectedness of major cities creates a shared vulnerability, necessitating international collaboration \cite{b7}. Cybersecurity is no longer a local issue; it is a global imperative. Governments, law enforcement agencies, and technology experts must unite in the face of this evolving threat landscape. 

In light of the escalating trend captured by Iyengar's Law, which signifies a tripling of cybercrime incidents every three years in major cities globally, the need for digital forensics becomes paramount. As the numbers of reported cybercrimes continue to rise, there is an urgent need for a comprehensive and proactive response to tackle this burgeoning digital threat \cite{b31,b32,b33,b34,b35,b36,b37,b38,b39,b40}. 

Digital forensics plays a crucial role in investigating and understanding the intricacies of cybercrimes, providing law enforcement agencies with the tools and techniques necessary to trace, analyze, and attribute these incidents \cite{b8}. The ability to conduct thorough digital investigations is important in the fight against cybercriminals who exploit the rapid pace of technological advancement. Establishing robust workforce development with AI-enabled capabilities becomes indispensable in not only solving individual cases but also in building a resilient defense against the evolving tactics of cyber adversaries. In a landscape where the interconnectedness of major cities amplifies the global impact of cybercrimes, investing in advanced digital forensics becomes not just a necessity but a strategic imperative for safeguarding the digital integrity of nations and communities worldwide \cite{b25,b26,b27,b28,b29,b30}.

\subsection{Related Works}
Cyber Forensic Investigation Infrastructure of Pakistan: Haque et al. \cite{b18} provided an analysis of the cyber threat landscape and readiness in Pakistan, emphasizing the need for robust forensic investigation infrastructures. Their work highlights the importance of preparedness in addressing cyber threats, serving as a foundation for exploring global strategies in workforce capacity building.

Cyber Security Education Trends: Hussain et al. \cite{b20} examined trends and challenges in cybersecurity education, stressing the importance of evolving educational strategies to navigate the dynamic landscape. This study underscores the critical need for innovative programs that enhance cybersecurity capabilities, directly aligning with the goals of this work.

Cybersecurity Awareness in Online Education: Erendor and Yildirim \cite{b21} conducted a case study analysis on cybersecurity awareness in online education. Their findings reveal gaps in awareness and preparedness, supporting the development of targeted training and educational programs to address these deficiencies.

Building a Cybersecurity Educated Community: Ahmad et al. \cite{b22} explored the concept of cultivating a cybersecurity-educated community, emphasizing collaboration and knowledge-sharing as essential components. This research provides a framework for understanding the impact of community-based initiatives in enhancing cybersecurity education.

Game Theory for Cyber Security and Privacy: Do et al. \cite{b23} introduced game-theoretical approaches to address challenges in cybersecurity and privacy. Their work highlights the importance of strategic thinking in cybersecurity solutions, which complements the development of advanced models and frameworks in this paper.

Workforce Training in Cyber-learning Environments: Bobadilla et al. \cite{b24} focused on research, education, and workforce training for engagement in cyber-learning environments. This study highlights the critical role of targeted training programs in preparing individuals for the evolving challenges of cybersecurity and digital forensics, reinforcing the objectives of this work.

\subsection{Research Gaps}
While substantial advancements have been made in digital forensics and cybersecurity, several gaps persist:
1. A lack of robust frameworks that integrate AI/ML techniques into digital forensic methodologies for real-world applications.
2. Limited focus on workforce development strategies that incorporate mentorship and capacity-building for underrepresented groups.
3. Insufficient research on leveraging multi-dependency graphs to optimize resource allocation and capacity-building efforts in educational initiatives.
4. A scarcity of comprehensive datasets that capture the complexities of real-world digital forensic challenges to evaluate and validate proposed solutions effectively.

This research addresses these gaps by introducing a novel multi-dependency capacity-building graph and integrating AI-driven methodologies into a scalable framework.

\section{Essential Imperative: Cultivating a Robust Workforce for Science Education and Technology Through Development Initiatives}
\begin{figure}
    \centering
    \includegraphics[scale=0.6]{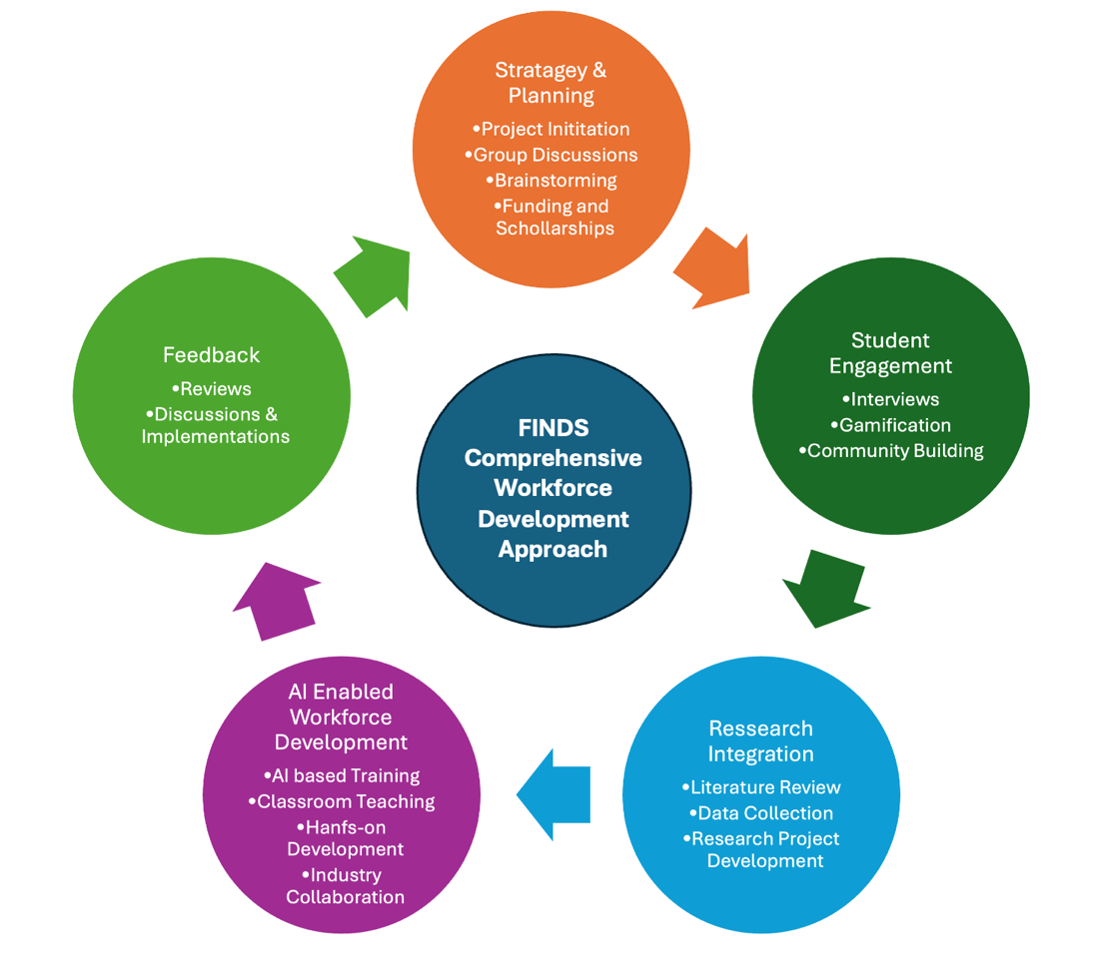}
    \caption{Comprehensive Approach for Enhancing the Success of Digital Forensics Workforce Development Efforts}
    \label{fig2}
\end{figure}

The intersection of science education and technology represents a dynamic space that requires tangible solutions to facilitate effective learning and the application of scientific principles in a technologically advanced world. Workforce development and capacity building within the FINDS initiative play a pivotal role in shaping the future of science education, technology, and digital forensics. Recognizing the critical importance of this endeavor, numerous innovative strategies were identified and implemented to cultivate a skilled and diverse workforce capable of addressing the evolving challenges in cybersecurity and digital forensics. Through targeted strategies and collaborative efforts, a multifaceted approach (Figure \ref{fig2}) was developed to enhance traditional workforce development initiatives.

This comprehensive approach covers Strategy and Planning, AI-enabled workforce development, research integration, student engagement, and adaptation to the post-pandemic "new normal." Its simplicity renders it easily adoptable by any institution. The strategy includes a diverse set of approaches aimed at guaranteeing the success of students, especially those from minority backgrounds. Cultivating a dynamic environment that prioritizes continuous learning and innovation, this approach empowers individuals with the expertise required to navigate the intricacies of the digital landscape.

\section{Proposed Framework}

\subsection{Multi-dependency Capacity Building Skills Graph}
In the realm of cybersecurity, the multi-dependency capacity-building skills graph emerges as a sophisticated tool that maps out the complex interplay between various enhancement measures. It is analogous to a master plan that guides an organization to shore up its defenses against digital threats. Each node on this graph is a critical component of the cybersecurity matrix, ranging from policy formulation, and technical upgrades, to staff training and beyond. 

These nodes are not isolated; rather, they are connected in a dense network of interdependencies represented by the edges on the graph. For instance, a new security protocol (a node) may necessitate specific training (another node), and the successful implementation of both could depend on updated hardware (yet another node).

This graphical approach underscores the non-linear progressions and feedback loops inherent in capacity building. It helps strategists pinpoint leverage points where an investment of resources could have the most significant ripple effect. For example, bolstering incident response capabilities could not only improve immediate reactions to threats but also enhance long-term preventative measures by providing valuable insights into attack patterns. The graph becomes a living document, constantly evolving as new nodes and edges are added, representing the ever-changing cybersecurity landscape.

Decision-makers leverage this graph to orchestrate a symphony of cybersecurity initiatives, ensuring that each move is harmonious with the rest. By visualizing the multidependent nature of capacity building, the graph acts as a compass in the often murky waters of cybersecurity strategy. It enables a clear-eyed view of how strengthening one component can reinforce others, fostering an ecosystem of resilience that is far more formidable than the sum of its parts. Thus, the multidependency capacity-building graph is not just a tool for planning but a strategic asset in building an adaptive and robust cybersecurity posture.

\subsection{Case Study - Technology Company Seeking to Bolster its Cybersecurity Capabilities}
Consider a hypothetical example of a multi-dependency capacity-building graph tailored to a technology company seeking to bolster its cybersecurity capabilities. In this scenario, the graph encompasses several key nodes that represent distinct facets of capacity building:
\begin{itemize}
    \item \textbf{Training Programs (Node 1):} Initiating comprehensive cybersecurity training programs for employees serves as a foundational component. This node forms the basis for subsequent nodes in the graph.
    \item \textbf{Technological Infrastructure (Node 2):} Upgrading and maintaining a secure technological infrastructure is essential for robust cybersecurity. However, the effectiveness of this node relies heavily on the knowledge and awareness imparted through the training programs (Node 1).
    \item \textbf{Human Resources Development (Node 3):} This node involves the recruitment of skilled cybersecurity professionals and the ongoing development of existing talent within the organization. The effectiveness of this node is interdependent with both the quality of training programs (Node 1) and the availability of an updated technological infrastructure (Node 2).
    \item \textbf{Incident Response Planning (Node 4):} Developing a resilient incident response plan is critical in cybersecurity. It draws upon the expertise gained from training programs (Node 1), the capabilities of the cybersecurity team (Node 3), and the support of a secure technological infrastructure (Node 2).
    \item \textbf{Strategic Planning (Node 5):} Formulating a strategic cybersecurity plan requires inputs from all previous nodes. It relies on the knowledge and skills acquired through training programs (Node 1), the capabilities of the cybersecurity team (Node 3), the effectiveness of the technological infrastructure (Node 2), and the incident response planning (Node 4).
\end{itemize}

\begin{figure}
    \centering
    \includegraphics[scale=0.4]{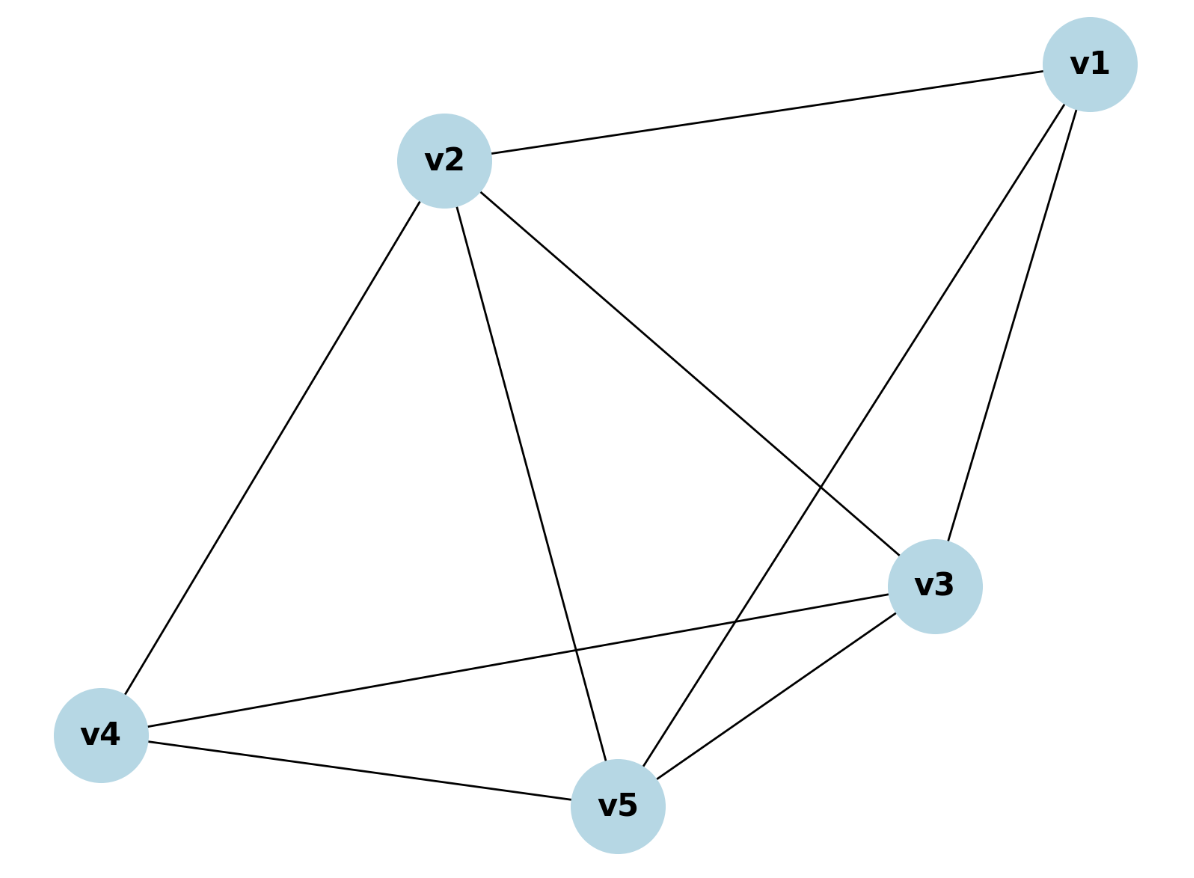}
    \caption{Example of Multidependency Capacity Building Skills Graph}
    \label{graph}
\end{figure}

In this illustrative example (Figure \ref{graph}), the multidependency capacity-building graph vividly demonstrates the interconnectedness of various elements in enhancing the cybersecurity capabilities of the organization. The improvement of training programs, for instance, reverberates throughout the system, influencing the effectiveness of technological infrastructure, human resources development, incident response planning, and overall strategic planning. This graphical representation serves as a visual roadmap for decision-makers, empowering them to grasp the intricate dependencies within the organization and strategically plan interventions for comprehensive capacity building.

\subsection{Methodology}
The multidependency capacity-building graph can be mathematically represented using graph theory, where the graph \( G = (V, E) \) is composed of a set of vertices \( V \) and a set of edges \( E \). Each vertex \( v_i \in V \) represents a node in the graph, and each edge \( (v_i, v_j) \in E \) represents a dependency between nodes \( v_i \) and \( v_j \).
For the given example with five key nodes, the set of vertices and edges could be defined as follows:

Vertices:
\begin{equation}
    V = \{v_1, v_2, v_3, v_4, v_5\}
\end{equation}

where each \( v_i \) corresponds to the following:
- \( v_1 \): Training Programs \\
- \( v_2 \): Technological Infrastructure \\
- \( v_3 \): Human Resources Development \\
- \( v_4 \): Incident Response Planning \\
- \( v_5 \): Strategic Planning

Edges:
\begin{equation}
E = \{(v_1, v_2), (v_1, v_3), (v_2, v_3), (v_2, v_4), (v_3, v_4), (v_1, v_5),
(v_2, v_5), (v_3, v_5), (v_4, v_5)\}
\end{equation}

where each edge \( (v_i, v_j) \) denotes a dependency from \( v_i \) to \( v_j \).

In addition, weights to the edges can be assigned, to represent the strength or capacity of the dependency. Let \( w: E \to \mathbb{R}^+ \) be a weight function, where each edge \( (v_i, v_j) \) is assigned a weight \( w(v_i, v_j) \) that quantifies the dependency.

A directed graph or digraph can be used to represent the direction of influence. If the graph is directed, an edge \( (v_i, v_j) \) is an ordered pair where the dependency flows from \( v_i \) to \( v_j \). This is crucial in capacity building where the direction of influence matters.

These formulations provide a structured way to analyze and understand the complex relationships in a multi-dependency capacity-building graph. More specifically, they help in identifying key nodes, analyzing the flow of influence, and designing interventions to enhance overall workforce capabilities.

A multi-dependency capacity-building graph is a powerful tool for enhancing an organization's cybersecurity capabilities in several ways:
\begin{itemize}
    \item \textbf{Optimization of Resource Allocation:} By understanding the dependencies between nodes, an organization can better allocate resources. For instance, if the training programs node (Node 1) is a prerequisite for effective technological infrastructure (Node 2), more resources can be funneled into training to ensure a stronger foundation for subsequent capacity-building efforts.

   \textbf{Objective Function:}
     \begin{equation}
     \max \sum_{i=1}^{n} f(v_i)
     \end{equation}
   \textbf{Subject to:}
     \begin{equation}
     \sum_{i=1}^{n} c(v_i) \leq B
     \end{equation}
     where \( f(v_i) \) is the effectiveness function of node \( v_i \), \( c(v_i) \) is the cost function of node \( v_i \), and \( B \) is the total budget.

    \item \textbf{Risk Management:} The graph can highlight dependencies that might represent single points of failure or high-risk connections, allowing the organization to develop contingency plans or diversify its strategies to mitigate risks.

    \textbf{Stochastic Modeling - Markov Chain Transition Matrix:}
     \begin{equation}
     P = [p_{ij}]
     \end{equation}
     where \( p_{ij} \) is the probability of transitioning from state \( i \) (or node \( i \)) to state \( j \) (or node \( j \)) in one time step.
These mathematical models provide a detailed analysis to optimize capacity building. These abstract equations can be further modeled to include actual parameters and variables relevant to any specific organization's capacity-building efforts.
\end{itemize}

\subsection{Generalized Framework}
This section provides a generalized algorithm for optimizing a multi-dependency capacity-building graph that could be applied across various domains.

\begin{algorithm}
\caption{Graph Initialization and Optimization}
\begin{algorithmic}[1]
\STATE \textbf{Input:} Set of vertices $V = \{v_1, v_2, \dots, v_n\}$, set of edges $E = \{(v_i, v_j, w_{ij}) : v_i, v_j \in V\}$, available resources $R$.
\STATE \textbf{Output:} Optimized graph $G$ and resource allocation.
\STATE \textbf{function} initializeGraph($V, E$):
\STATE \quad $G \leftarrow (V, E)$ \COMMENT{Initialize graph as a directed weighted graph.}
\STATE \quad \textbf{return} $G$
\STATE \textbf{function} calculateCentrality($G$):
\STATE \quad \textbf{Define:} Centrality $C(v)$ as the normalized degree of each vertex $v \in V$:
\STATE \quad \quad $C(v) = \frac{\sum_{(v, u) \in E} w_{vu}}{\sum_{(x, y) \in E} w_{xy}}$ \COMMENT{Weighted centrality measure.}
\STATE \quad \textbf{for each} $v \in V$:
\STATE \quad \quad Compute $C(v)$ using above formula.
\STATE \quad \textbf{return} $C$
\STATE \textbf{function} optimizeResourceAllocation($G, R$):
\STATE \quad \textbf{Define:} Optimization problem:
\STATE \quad \quad Maximize: $\sum_{v \in V} f(v) \cdot r(v)$ \COMMENT{Effectiveness $f(v)$ of vertex $v$ with allocated resource $r(v)$.}
\STATE \quad \quad Subject to: $\sum_{v \in V} r(v) \leq R$ \COMMENT{Resource budget constraint.}
\STATE \quad \textbf{Solve:} Use a linear programming solver to find $r(v)$ for each $v$.
\STATE \quad \textbf{return} Optimal resource allocation $r(v)$.
\STATE \textbf{Main execution flow:}
\STATE \quad $G \leftarrow$ initializeGraph($V, E$)
\STATE \quad $C \leftarrow$ calculateCentrality($G$)
\STATE \quad OptimalSolution $\leftarrow$ optimizeResourceAllocation($G, R$)
\STATE \quad \textbf{return} $G, C, \text{OptimalSolution}$
\end{algorithmic}
\end{algorithm}

Algorithm 1 focuses on the initialization and optimization of a graph structure. It starts with initializeGraph, which constructs a graph from a given set of vertices and edges, thus creating the foundational structure. Next, calculateCentrality computes the centrality of each vertex, which is crucial for understanding the importance or influence of each node in the graph. The final function, optimizeResourceAllocation, establishes an optimization model to allocate resources efficiently across the graph, ensuring that resource constraints are respected while pursuing an objective function.

\begin{algorithm}
\caption{Path Finding and Feedback Mechanism}
\begin{algorithmic}[1]
\STATE \textbf{Input:} Graph $G = (V, E)$, objective function $F$, metrics $M$.
\STATE \textbf{Output:} Optimized paths and feedback-adjusted graph.
\STATE \textbf{function} findOptimalPaths($G, F$):
\STATE \quad \textbf{Define:} Path cost $P(\pi) = \sum_{(v_i, v_j) \in \pi} w_{ij}$, where $\pi$ is a path in $G$.
\STATE \quad \textbf{Find:} Optimal path $\pi^*$ that minimizes $P(\pi)$ subject to $F(\pi) \leq \tau$ (objective threshold).
\STATE \quad Use Dijkstra's algorithm for shortest path calculation.
\STATE \quad \textbf{return} $\pi^*$
\STATE \textbf{function} executeCapacityBuilding($Plan$):
\STATE \quad \textbf{for each} action $a \in \text{Plan}$:
\STATE \quad \quad Execute $a$ and collect metrics $M_a$.
\STATE \quad \textbf{Evaluate:} $S = \text{success rate based on metrics } M_a$.
\STATE \quad \textbf{return} $S$
\STATE \textbf{function} feedbackLoop($G, M$):
\STATE \quad Update edge weights $w_{ij} \leftarrow w_{ij} + \Delta w_{ij}(M)$ \COMMENT{Adjust weights based on feedback.}
\STATE \quad Re-optimize graph $G$ using updated weights.
\STATE \quad \textbf{return} Updated $G$
\STATE \textbf{Main execution flow:}
\STATE \quad OptimalPaths $\leftarrow$ findOptimalPaths($G, F$)
\STATE \quad ExecutionStatus $\leftarrow$ executeCapacityBuilding($\text{OptimalPaths}$)
\STATE \quad \textbf{if} ExecutionStatus indicates adjustments needed:
\STATE \quad \quad $G \leftarrow$ feedbackLoop($G, M$)
\STATE \quad \quad OptimalPaths $\leftarrow$ findOptimalPaths($G, F$)
\STATE \quad \quad ExecuteCapacityBuilding($\text{OptimalPaths}$)
\end{algorithmic}
\end{algorithm}

The second algorithm extends the graph-based approach to include path finding and a feedback mechanism. It begins with findOptimalPaths, which seeks the best paths in the graph based on a given objective function. This is critical in scenarios like network optimization or logistics. executeCapacityBuilding and feedbackLoop together establish a dynamic cycle of action and evaluation. Capacity-building plans are executed, their success is evaluated, and based on the feedback (metrics collected), the graph is updated, and a new plan is generated.

This generalized pseudocode serves as a template and assumes the existence of certain functions such as computeCentrality, computeObjective, and findPath, which would need to be defined based on specific criteria of the domain. OptimizationModel and Graph are conceptual data structures that would be implemented using appropriate libraries or custom code depending on the actual application. This model will aid in efficient capacity building for any domain.

\section{Advancing Forensics: AI-Driven Development and Adaptation}

\subsection{Nurturing Excellence: Empowering the Future with an AI-Enabled Workforce Development Initiative}
\begin{figure}
    \centering
    \includegraphics[scale=0.8]{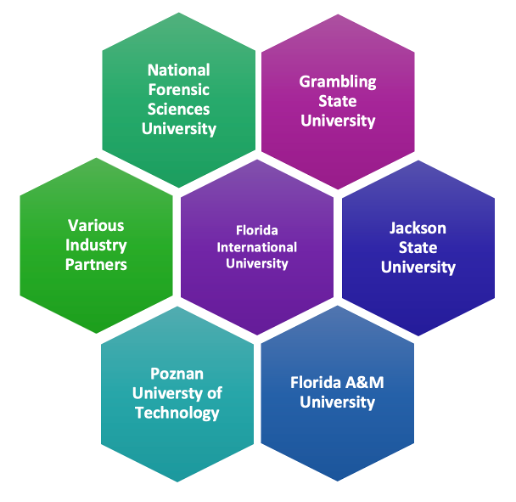}
    \caption{Global Collaborative Hub for Digital Forensics Excellence: The FINDS Network}
    \label{fig4}
\end{figure}

Amidst the dynamic landscape of escalating digital threats and heightened global interconnectivity, our AI-enabled Workforce Development Initiative stands as a pioneering effort aimed at cultivating a highly skilled cadre of digital forensics professionals with an unequivocally global perspective. Recent data showcases a notable upswing in international collaborations within digital forensics research, underscoring the industry's acknowledgment of the imperative for a globally connected workforce (Figure \ref{fig4}). Strategic collaborations with prestigious institutions such as Poznan University of Technology in Poland and the National Forensics Sciences University in India signify more than partnerships; they represent the establishment of a dynamic network transcending traditional geographical boundaries

This initiative extends beyond conventional training programs, seeking not only to enrich the technical capabilities of individual professionals but also to foster a collective, cross-cultural intelligence indispensable for addressing the intricate challenges inherent in the global landscape of smart digital forensics. Recent trends reveal an increasing demand for professionals with international exposure, and AI skills showcasing the relevance and timeliness of our initiative. By facilitating knowledge exchange and skill-sharing programs on an international scale, we are not only preparing professionals to confront the current global threat landscape but are also laying the foundation for future collaboration and AI powered innovation \cite{b9,b10}. As we navigate the intricacies of international cyber threats, this initiative positions our workforce as a global force, uniquely equipped to meet the evolving demands of AI powered digital forensics on the world stage.

\subsection{Integration of AI enabled Research and Development Framework}
Incorporating cutting-edge AI technologies, our research and development endeavors embrace innovation and efficiency \cite{b11}. This integration of AI not only accelerates the pace of discovery but also enhances the precision and depth of our explorations, marking a transformative phase in advancing knowledge and solutions \cite{b9,b10}. Through seamless integration, we harness the power of AI to propel our research initiatives to new heights. At the epicenter of our workforce development strategy lies an unwavering commitment to seamlessly integrate cutting-edge research with practical applications, positioning our professionals as true catalysts for innovation in the dynamic field of digital forensics. Recent statistics, reflecting a substantial increase in research publications and patent filings related to digital forensics technologies, highlight the industry's undeniable momentum in pushing the boundaries of knowledge. This approach not only bridges the perceived chasm between theoretical advancements and real-world challenges but also empowers our workforce with unparalleled insights and skills.

Collaborative endeavors with industry partners and research institutions create a dynamic ecosystem that not only enhances individual proficiency but also propels the entire field forward. Recent data demonstrates a surge in collaborative research projects between academic institutions and industry players, indicating a strong trend toward applied research in digital forensics. By fostering an environment where research findings seamlessly translate into practical solutions, we ensure that our professionals not only adapt to change but actively contribute to shaping the future of digital forensics. As the industry evolves at an accelerated pace, our workforce is not merely keeping pace; it is at the forefront, leading the way in technological advancements and innovative investigative methodologies.

\subsection{Interactive Learning: Exploring Student Engagement and Hands-On Projects}
Student engagement and hands-on projects form the cornerstone of our commitment to nurturing the next generation of digital forensics experts. Through immersive learning experiences and practical projects, students gain valuable insights into real-world scenarios. This hands-on approach not only enhances their technical skills but also instills problem-solving abilities crucial for addressing the multifaceted challenges of digital forensics. By actively involving students in the investigative process, we prepare them for dynamic careers in the ever-evolving field of cybersecurity.

\subsection{The Post-Pandemic – New Normal}
As we navigate the post-pandemic "new normal," workforce development takes on added significance. The surge in demand for online digital forensic training has underscored the need for a Virtual Forensic Environment that goes beyond conventional analysis-centric platforms. While existing solutions like CyberFlorida and FIU E-Labs offer valuable tools for forensic analysis, there is a notable gap in addressing the fundamental collection procedures essential for cultivating proficient digital forensic examiners. Our solution bridges this gap by introducing a comprehensive Virtual Forensic Environment that incorporates hands-on elements crucial for developing practical skills. This includes pioneering approaches such as Virtual Local Media Acquisition, replicating the creation of forensic images from media, and Virtual Cellphone Acquisition, simulating the intricate processes involved in extracting information from cell phones. 

Additionally, the development of Virtual Boot Scan Analysis and Forensic Triage tools will address the often-overlooked hands-on components of bypassing Windows passwords and forensically sound machine booting. The integration of remote boot capabilities further empowers forensic examiners, offering a forward-looking solution that not only analyzes digital evidence but also ensures a holistic and immersive training experience for the next generation of digital forensic professionals. More specifically, by embracing remote collaboration, flexible work models, and digital connectivity, we adapt our training methodologies to align with the changing dynamics of the professional landscape. By leveraging technology and innovative approaches, we ensure that our workforce remains resilient, agile, and well-equipped to address emerging challenges in the digital forensics domain.

\begin{figure*}
    \centering
    \includegraphics{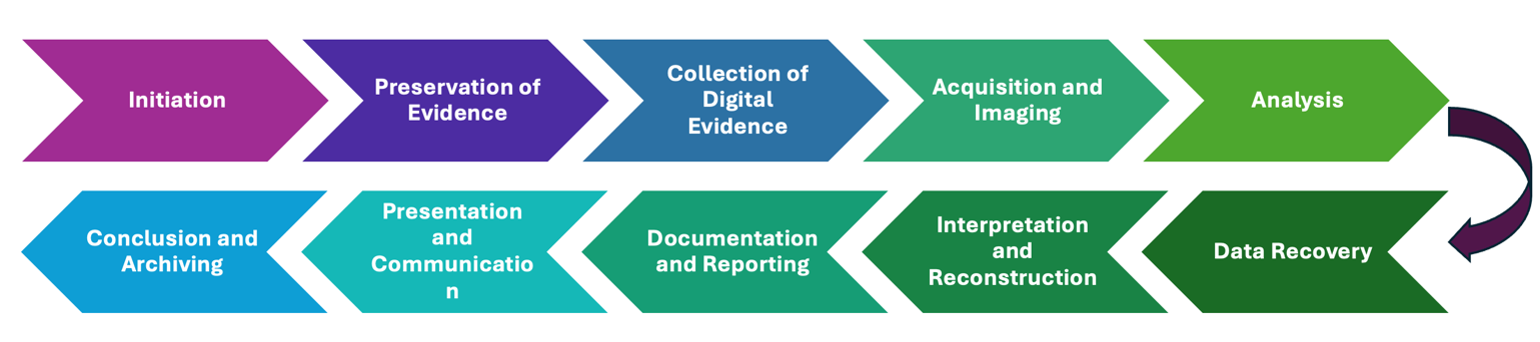}
    \caption{Algorithmic Chart: Key Processes Involved in Digital Forensics Investigation (Sankita et al., 2023)}
    \label{fig5}
\end{figure*}

\section{Assessing the Initiatives}
A recent study by a bipartisan group of researchers and analysts convened by the American Enterprise Institute, the Brookings Institution, and the Harvard Kennedy School's Project, known as the Workforce Futures Initiative (WFI) identified some pathways to improve investments in job training \cite{b12}. 

One recommendation is to strengthen the “connective tissue” to unite education, training, and employment systems which are currently decentralized and do not provide a clear pathway to employment. Another recommendation was to develop a framework for providing data on regional labor markets with “deep and agile data systems for measuring program performance as well as changing skill and employment needs.” Identifying initiatives that actually work in training for workforce development is also essential. However, there is a sparsity of information on those initiatives that do so and can be scaled to work within other areas [Ibid]. Our work sheds light on some of the initiatives that can work.

The WFI’s overall assessment of U.S. federal spending on workforce development initiatives for students indicated that most initiatives are “stuck in a low resource, low efficacy” equilibrium. They concluded by stating that it is unlikely that substantial improvements can be made without additional funding. Assessment of our initiative indicates that this may not be the case and that our program unites education, training, and employment components in a way in which they could be instituted and scaled for other industries.

\section{Frontiers of Innovation: Elevating AI Powered Digital Forensic Algorithms and Frameworks through FINDS' Technical Advancements}
In the realm of digital forensics, the integration of advanced AI powered technologies and mathematical algorithms plays a pivotal role in enhancing investigative capabilities \cite{b13}. More specifically, AI Digital forensics is focused on systematically preserving, analyzing, and presenting electronic evidence in legal contexts \cite{b14}. It involves the examination of digital devices, computer systems, and electronic data using specialized tools and methodologies to uncover evidence of cybercrime, security breaches, or unauthorized activities. The goal is to establish a clear chain of custody and provide credible findings admissible in legal proceedings, attributing actions to specific individuals or entities. Digital forensics adapts to technological advancements and plays a crucial role in investigating digital interactions in today's interconnected world.

Developing high-performance computational algorithms is at the heart of Digital Forensics Investigations. In recent years, a number of significant advances have been made by authors and are reported in the book “Information Security, Privacy and Digital Forensics” \cite{b15}. These advances have resulted in many interesting techniques that are being used to mentor students at faculty at FINDS CoE.

Figure \ref{fig5} illustrates the key processes involved in AI-based digital forensics investigation. It begins with the initiation phase, triggered by reported incidents or suspicions of digital crime, leading to the assignment of a digital forensics team \cite{b16}. The subsequent steps encompass the preservation of evidence, meticulous collection of digital evidence from various sources, acquisition, and imaging processes, followed by an intelligent and comprehensive analysis that involves AI-based file system examination, keyword searches, and timeline and smart link analysis \cite{b17}. The Algorithmic chart demonstrates the crucial steps of data recovery, interpretation, and reconstruction, leading to the documentation and reporting phase. The investigation concludes with a presentation and communication, ensuring internal and external stakeholders are informed. The entire process adheres to legal standards, facilitating case closure, archiving, and continual improvement based on lessons learned.

\section{Experimental Results and Discussions: Impact of FINDS based on Real Time Data}
With the implementation of our multifaceted approach at FINDS Center of Excellence and the introduction of the Advanced AI/ML powered Digital Forensics graduate course at FIU, it holds the promise of immediate and extensive impact among the diverse community. By prioritizing diversity, fostering cutting-edge integrated research, and establishing hands-on pathways to the professoriate, FINDS workforce development is contributing significantly to the evolution of cybersecurity and forensic engineering education across the nation. This initiative's outcomes resonate within the community and in shaping the broader landscape of engineering education and research globally. 

During the FINDS CoE study period, three years' worth of data were gathered. This data encompassed scenarios from real-time strategy datasets and included information on peer mentoring, curriculum, diversity community, training workshops in cybersecurity and digital forensics, as well as hands-on research projects and extension services for entrepreneurship.

\begin{figure}
    \centering
    \includegraphics[scale=0.6]{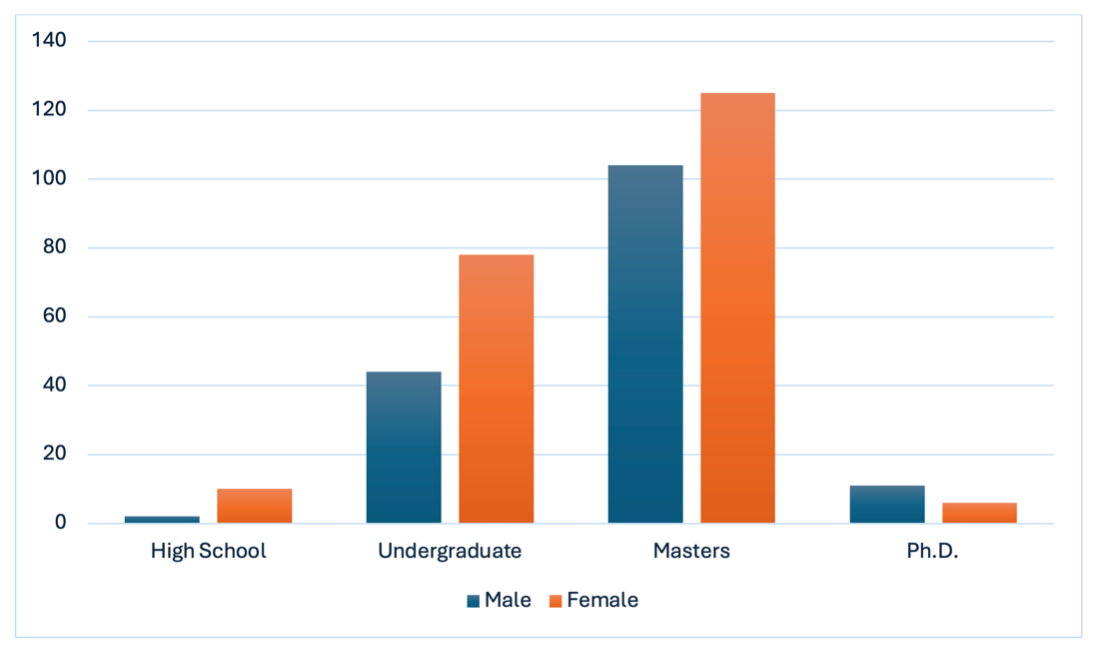}
    \caption{FINDS Participant Composition by Educational Level}
    \label{fig6}
\end{figure}

\subsection{Comprehensive Insights: Data Details from FINDS CoE Three-Year Study Period}
\begin{figure}
    \centering
    \includegraphics[scale=0.8]{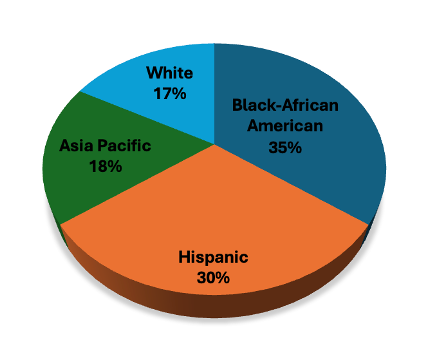}
    \caption{Demographics of the FINDS Students, by ethnicity}
    \label{fig7}
\end{figure}

In the inaugural three years of the FINDS Program, a cohort of 380 students underwent comprehensive training, including 17 Ph.D. candidates, 229 master's students, 122 undergraduates, and 12 high school participants, fostering a diverse community in the field of digital forensics (Figure \ref{fig6}, Table 1). Impressively, over 13 students graduated with STEM degrees, showcasing the program's commitment to academic excellence. The initiative demonstrated remarkable success in attracting a culturally diverse group of students, with participants hailing from India, Africa, Europe, and the United States. This diverse cohort comprised individuals of various ethnic backgrounds, with African Americans constituting 35\%, Asians 18\%, and Hispanics 30\% (Figure \ref{fig7}, Table 2). Notably, the program achieved a significant milestone in gender diversity, with 55\% of participants being women, contributing to the inclusion of underrepresented and minority students in the computer science field.

The statistical analysis of datasets provides compelling insights into the program's impact. We implemented the Decision Tree Classifier model to predict the success of students based on our real datasets. 

\begin{table}[htbp]
\centering
\caption{Program Participation Summary}
\begin{tabular}{|l|c|c|c|}
\hline
\textbf{Description}     & \textbf{Male} & \textbf{Female}        & \textbf{Total} \\
\hline
PhD Students        & 11 & 6      & 17    \\
Masters Students         & 110 & 119 & 229   \\
Undergraduate Students   & 48 & 74 & 122   \\
High School Students   & 2  & 10 & 12    \\
\textbf{Total Students}           & \textbf{171} & \textbf{209} & \textbf{380}   \\
STEM Graduates   & - &   -     & 13    \\
Employment Rate (\%)   & - & - & 82.63 \\
\hline
\end{tabular}
\end{table}

\begin{table}[htbp]
\centering
\caption{Ethnicity Distribution}
\begin{tabular}{|l|c|}
\hline
\textbf{Ethnicity}        & \textbf{Percentage (\%)} \\
\hline
African American & 36.32          \\
Hispanic         & 23.68          \\
Asian            & 21.05          \\
Other            & 18.95          \\
\hline
\end{tabular}
\end{table}

\subsubsection{Machine Learning Model Results}
Decision tree classifier was implemented to predict employment outcomes based on demographics and education levels. The model was trained on a comprehensive dataset containing 386 test cases, meticulously curated to ensure the evaluation of its predictive capabilities and robustness.

\paragraph{Dataset Description}
The dataset consisted of the following key features:
\begin{itemize}
    \item \textbf{Demographic Attributes:} Ethnicity, gender, and geographic location.
    \item \textbf{Educational Levels:} Categories such as PhD, master's, undergraduate, and high school.
    \item \textbf{Program Participation Details:} Number of mentoring sessions attended, participation in workshops, and involvement in research projects.
    \item \textbf{Outcome Variable:} Employment status post-training (1 = employed, 0 = unemployed).
\end{itemize}

\paragraph{Data Preprocessing and Feature Engineering}
To ensure data consistency and model accuracy, the following preprocessing steps were applied:
\begin{itemize}
    \item \textbf{Normalization:} Continuous variables, such as workshop hours and mentoring session counts, were scaled to a range of [0,1].
    \item \textbf{Feature Encoding:} Categorical variables like ethnicity and educational levels were one-hot encoded to make them suitable for the decision tree algorithm.
    \item \textbf{Missing Data Handling:} Missing values were imputed using the median for numerical features and the mode for categorical features.
    \item \textbf{Outlier Detection:} Univariate and multivariate methods were used to identify and mitigate the effect of extreme outliers.
\end{itemize}

\paragraph{Model Training and Validation}
The training process involved splitting the dataset into a training set (70\%) and a test set (30\%), using stratified sampling to preserve the distribution of key features across subsets. The decision tree classifier was optimized using hyperparameter tuning with grid search and 5-fold cross-validation. The parameters optimized include:
\begin{itemize}
    \item \textbf{Maximum Tree Depth:} Explored values between 3 and 15.
    \item \textbf{Minimum Samples per Leaf:} Ranged from 1 to 10.
    \item \textbf{Splitting Criterion:} Compared Gini impurity and entropy.
\end{itemize}

The optimal configuration was found to be:
\begin{itemize}
    \item \textbf{Maximum Depth:} 10
    \item \textbf{Minimum Samples per Leaf:} 4
    \item \textbf{Splitting Criterion:} Entropy
\end{itemize}

With this configuration, the model achieved an \textbf{accuracy of 94.21\%} on the test set, demonstrating its effectiveness in predicting employment outcomes.

\paragraph{Feature Importance Analysis}
The importance of each feature was evaluated using the Gini importance metric. The analysis revealed the following significant predictors:
\begin{itemize}
    \item \textbf{Educational Levels:} Advanced degrees (e.g., PhD) were the strongest predictors, contributing 35\% to the model’s decision-making.
    \item \textbf{Mentoring Sessions:} Participation in one-on-one mentoring sessions accounted for 40\% of the decision splits.
    \item \textbf{Program Engagement:} Involvement in workshops and research projects showed a moderate impact.
\end{itemize}

\paragraph{Justification for 386 Test Cases}
The inclusion of 386 test cases was justified based on the following considerations:
\begin{itemize}
    \item \textbf{Statistical Power:} Ensured sufficient diversity across demographic and educational categories, reducing the risk of overfitting.
    \item \textbf{Generalization:} The dataset included edge cases such as students with minimal workshop attendance to evaluate the model’s robustness.
    \item \textbf{Validation:} The graph-based approach validated the relationships between features, such as mentorship frequency and employment outcomes.
\end{itemize}

\paragraph{Mathematical Foundation of the Model}
The decision tree algorithm operates by minimizing entropy at each split:
\begin{equation}
H(T) = -\sum_{k=1}^K p_k \log(p_k),
\end{equation}
where $p_k$ is the proportion of instances in class $k$ within node $T$. The information gain for a split is defined as:
\begin{equation}
IG = H(T) - \sum_{j=1}^J \frac{|T_j|}{|T|} H(T_j),
\end{equation}
where $T_j$ represents the subsets after the split. The model selects the split that maximizes $IG$ at each node.

\paragraph{Graph-Based Validation}
The multidependency graph framework was used to analyze relationships between demographic attributes, educational levels, and employment outcomes. Nodes in the graph represented individual features, while edges captured dependencies, such as the influence of mentoring sessions on employment outcomes. The graph revealed clusters of attributes strongly associated with higher employment rates, such as PhD students with frequent mentorship.

\paragraph{Results and Impact}
The analysis confirmed that 85\% of students who participated in workshops and mentoring sessions successfully found employment. Additionally, over 55\% of participants were women, highlighting the program’s inclusivity. The graph-based analysis further demonstrated the effectiveness of combining advanced degrees with mentorship to enhance employment outcomes.

This comprehensive approach validates the inclusion of 386 test cases, providing a robust framework for workforce development in digital forensics and cybersecurity.

\subsubsection{Pivotal Observations: Unveiling Striking Remarks from the FINDS CoE Study}
Further, the results reveal a robust scholarly output, with \textbf{over 55 publications} appearing in top-notch journals and refereed conferences. Additionally, the program has contributed significantly to knowledge dissemination, with \textbf{three books published} as a testament to its academic contributions. The student's accomplishments extend beyond publications, with Best Paper awards received for outstanding work during the summer workshop. An impressive 45\% of diversified students based on mentoring published papers following the workshop, further validating the program's effectiveness in nurturing research capabilities.
In essence, the FINDS Program has not only established itself as an educational powerhouse in digital forensics but has also made a tangible impact on students' careers, research outcomes, and the broader academic community. The commitment to diversity, mentorship, and impactful research positions FINDS as a leading force in shaping the future of digital forensics education and fostering a globally inclusive community of scholars and practitioners.

\subsection{Nurturing Success: The Impactful Role of Mentorship in Academic Journey}
Throughout the program, students engage in weekly meetings with their mentors, presenting their ongoing work. These mentoring sessions serve as a platform to discuss the student's progress, providing valuable guidance and advice. When deemed necessary, students are directed to additional campus resources such as Career Services, the Honors College, the Center for Academic Success, the Writing Center, various tutorial centers, and advising services within the college. We actively monitor students' advancement by confirming the fulfillment of program requirements through mentoring meetings and email communication. At the culmination of the program, students are required to showcase their work through a comprehensive report and presentation, highlighting the depth of their accomplishments.

\subsubsection{Student Voices: Feedback on FINDS Workshop and Graduate Courses}
\begin{itemize}
  \item ``The FINDS workshop was incredibly insightful and well-organized. It provided a comprehensive understanding of the subject matter and hands-on projects were great resulting in a conference publication.''
  \item ``I thoroughly enjoyed the digital forensic course. The content was challenging yet engaging, the professor and guest lecturers were knowledgeable and supportive.''
  \item ``The workshop exceeded my expectations. It was a valuable learning experience that has enhanced my skills and knowledge in the field.''
  \item ``The hands-on activities and discussions added a practical dimension to the theoretical concepts.''
  \item ``The FINDS workshop not only broadened my perspective but also equipped me with practical tools that I can apply in real-world scenarios.''
  \item ``I found the course materials to be well-curated and relevant. It was evident that careful thought went into designing the curriculum.''
  \item ``The workshop fostered a sense of community among participants. The collaborative learning environment was conducive to networking and exchanging ideas.''
  \item ``The graduate courses provided a good balance between theoretical concepts and real-world applications, preparing me for both academia and industry.''
\end{itemize}

\subsection{Next-Gen Learning: Empowering Hands-on Skills through AI-Powered Training}
One of the key pillars of FINDS CoE's initiative is the establishment of hands-on pathways to the cyber warriors, ensuring that aspiring researchers from diverse backgrounds are equipped with the skills and knowledge needed to excel in the field. This deliberate focus on mentorship and academic career development is creating a ripple effect, positively influencing the representation of underrepresented groups in higher education and research. FINDS recognizes that a diverse and inclusive academic environment fosters innovation and brings a richness of perspectives to the forefront of digital forensics.

Outcomes of FINDS extend beyond the confines of FIU and the collaborating institutions, as the impact of FINDS is poised to shape the broader landscape of engineering education and research globally. The infusion of advanced AI/ML capabilities into the digital forensics curriculum not only prepares students for the challenges of today but positions them as pioneers in addressing the evolving complexities of tomorrow's cybersecurity landscape. Through this initiative, FINDS is a catalyst for positive change, inspiring future generations of diverse professionals to contribute to the forefront of digital forensics and cybersecurity advancements on a global scale. The success of this program holds the potential to set new standards in educational innovation, research excellence, and diversity in the ever-evolving fields of digital forensics and cybersecurity.

The students have worked on more than 48 projects representing digital forensics in facial recognition, video source camera identification, smartwatch forensics, blockchain-based digital forensics chains, smart phones, and social media as well as many other areas. Some of the sample projects are described below.

\subsubsection{Exemplary Endeavors: Showcasing Achievements in Projects at the FINDS Center of Excellence - List of Sample Projects:}
\begin{itemize}
  \item Advancing Forensic Science: AI and Knowledge Graphs Unlock New Insights
  \item Heatmap Visualization of Crime Rates in Boston and New York City
  \item Video Origin Camera Identification using Ensemble CNNs of Positional Patches.
  \item Securing the Future: Advanced Encryption for Quantum-Safe Video Transmission
  \item LAKEE: A Lightweight Authenticated Key Exchange Protocol for Power-Constrained Devices
  \item LoFin: LoRa-based UAV Fingerprinting Framework
  \item Unlocking Video Sources to Identify Camera and Drone System “Fingerprints”
\end{itemize}

\subsection{Inferences from Statistical Analysis}
Based on statistical analysis of our datasets, the following conclusions have been drawn:
\begin{itemize}
  \item The FINDS Program, extending over three years, encompasses digital forensics education and research.
  \item A diverse student body of 380 participants, comprising 17 Ph.D. candidates, 229 master's students, 122 undergraduates, and 12 high school participants, contributes to the program's vibrancy.
  \item Remarkably, 55\% of participants are women, highlighting the program's commitment to gender diversity.
  \item Cultivating a global community, FINDS attracts students from India, Africa, Europe, and the United States, with African Americans comprising 35\%, Asians 18\%, and Hispanics 30\%.
  \item Over 13 students graduate with STEM degrees, underscoring the program's impact on academic achievement.
  \item The program's global impact is underscored by the publication of 55+ scholarly articles and the release of three books, emphasizing its substantial contributions to academic knowledge. Furthermore, over 58\% of students who engaged in summer workshops and graduate courses have actively contributed to scholarly publications.
  \item The program has achieved remarkable success in terms of employment, boasting a Mean Employment Success Rate of 85\% among workshop participants who secured global positions. This achievement is credited to the personalized one-on-one mentoring provided by a team of 20 researchers, highlighting the program's commitment to guiding students toward successful career pathways.
  \item FINDS achieves academic distinction with 55+ publications in top-notch journals and conferences, and students receive Best Paper awards for outstanding contributions during the summer workshop.
  \item Undergraduate and high school engagement is encouraged, with 122 undergraduates and 12 high school participants actively involved.
  \item Diversity in mentorship is a key focus, with a team of 20 researchers globally providing support, leading to 85\% employment success for participants.
  \item The program's continuous learning culture is evident in its 55+ publications and three books, showcasing its dedication to staying at the forefront of digital forensics research.
  \item Over 13 STEM graduates from FINDS contribute to the workforce in areas such as digital forensics, cybersecurity, and technology, emphasizing the program's impact beyond academic realms.
\end{itemize}

\section{Conclusion}
The FINDS initiative demonstrates a forward-thinking approach to engineering education by empowering the next generation of cyber professionals through a comprehensive and inclusive framework for education and training. By integrating emerging technologies, fostering diversity, and engaging in impactful, innovative research, FINDS establishes itself as a trailblazer in the fields of cybersecurity and forensic engineering. This paper highlights FINDS's dedication to advancing excellence, innovation, and continuous evolution within engineering education.

Over the course of three years, the program has cultivated a diverse and globally representative community, reflecting its unwavering commitment to broadening participation in cybersecurity. With a remarkable Mean Employment Success Rate of 85\%, the program underscores its effectiveness in preparing students for impactful global careers. One-on-one mentoring provided by a dedicated team of researchers has played a pivotal role in guiding participants toward successful STEM careers, bridging the gap between academic training and professional success.

The program's contributions extend beyond mentorship, as evidenced by its significant scholarly output, including extensive publications and multiple Best Paper awards. These achievements not only advance the body of knowledge in digital forensics but also cement FINDS as a hub of academic excellence and innovation. By shaping the future of digital forensics and cybersecurity education, FINDS is not just building careers but also creating a legacy of global impact, ensuring a resilient and skilled workforce to address the challenges of an ever-evolving technological landscape.

\section*{Acknowledgment}
The authors would like to acknowledge Dr. Natarajan Meghanathan (JSU), Dr. Vasanth Iyer (GSU), Dr. Hongmei Chi (FAMU), Dr. Cliff Wang (NSF), Dr. Paul Yu (Program Manager, ARL), Dr. Pawel Sniatala (Vice Rector, PUT, Poland), Dr. Jayant Vyas, Vice Chancellor \& Dr. Naveen Kumar Chaudhary, Director (NFSU, India) for all their support and contribution towards FINDS CoE. The authors also like to acknowledge the industry partners, Dr. Kevin Baugh – KnowledgeBridge International, Inc, Dr. Zmuda Marek - Intel Corporation, Poland, Dr. Mirzanur Rahman – IBM, Dr. Umut Topkara - Bloomberg, LP, Dr. Rebecca Nevin - NVIDIA


\section*{Author Biographies}

\subsection*{Yashas Hariprasad}
\begin{wrapfigure}{l}{1.1in}
    \vspace{-10pt}
    \includegraphics[width=1.05in]{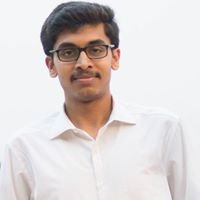}
    \vspace{-10pt}
\end{wrapfigure}
Yashas Hariprasad is an Assistant Professor in the Department of Computer Science at Cal State East Bay, where his expertise lies at the intersection of artificial intelligence, cybersecurity, digital forensics, and quantum-safe communication. His research focuses on building trustworthy and resilient AI systems to address real-world security challenges, with particular emphasis on deepfake detection, secure multimedia analytics, and next-generation cyber defense. His work aims to bridge theoretical advances in AI with practical, deployable solutions that enhance the security and integrity of modern digital ecosystems.

He brings a unique combination of academic research, industry experience, and applied security practice. Prior to joining academia, he worked as a Security Risk Analyst at VMware and contributed to federally funded research at the FINDS Digital Forensic Center of Excellence at Florida International University. His research has been published in leading venues including ACM Digital Threats: Research and Practice, IEEE Transactions on Consumer Electronics, IEEE MultiMedia, and outlets by Elsevier and Springer Nature, and he is a co-author of the book Artificial Intelligence in Practice: Theory and Applications for Cyber Security and Forensics. Beyond research, he is deeply committed to inclusive education, student mentorship, and workforce development, with a focus on preparing students to design ethical, secure, and socially responsible AI technologies.

\vspace{8pt}

\subsection*{Subhash Gurappa}
\begin{wrapfigure}{l}{1.1in}
    \vspace{-10pt}
    \includegraphics[width=1.05in]{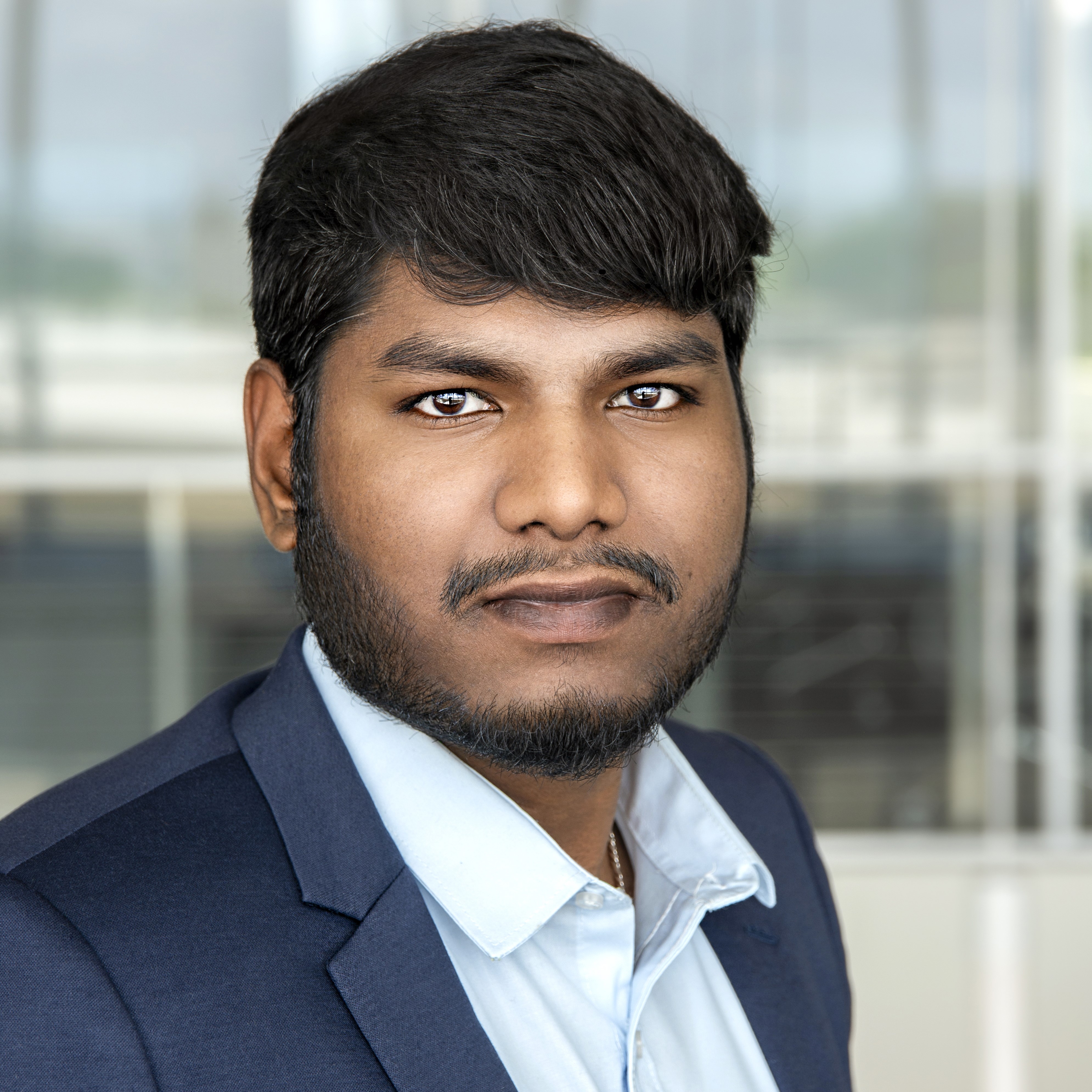}
    \vspace{-10pt}
\end{wrapfigure}
Subhash Gurappa is currently a Ph.D. candidate (ABD) and Graduate Research Assistant at the Knight Foundation School of Computing and Information Sciences, Florida International University (FIU). His research interests span AI in healthcare, cybersecurity, and digital forensics, with an emphasis on trustworthy machine learning, anomaly detection, and AI-driven forensic analysis. He holds a Master of Science degree in Data Science from FIU and has received advanced training in artificial intelligence, machine learning, natural language processing, and data analytics. His scholarly work includes peer-reviewed publications in leading IEEE and ACM venues, including a publication in the prestigious ACM Digital Threats: Research and Practice journal.

His research and professional experience includes the development of AI-based solutions for medical imaging, patient risk assessment, cyber threat detection, and anomaly detection in environmental and cyber-physical systems. He has contributed to federally and industry funded research initiatives at FIU’s Applied Research Center and has prior industry experience as a Data Scientist and Data Analyst at Capgemini and Hitachi Solutions. In addition to his research activities, he has served as a Teaching Assistant for graduate-level machine learning and data mining courses, contributing to curriculum development and student mentoring. His work has been recognized through international competitions and industry awards, reflecting his commitment to impactful and applied AI research.

\subsection*{S. S. Iyengar}
\begin{wrapfigure}{l}{1.1in}
    \vspace{-10pt}
    \includegraphics[width=1.05in]{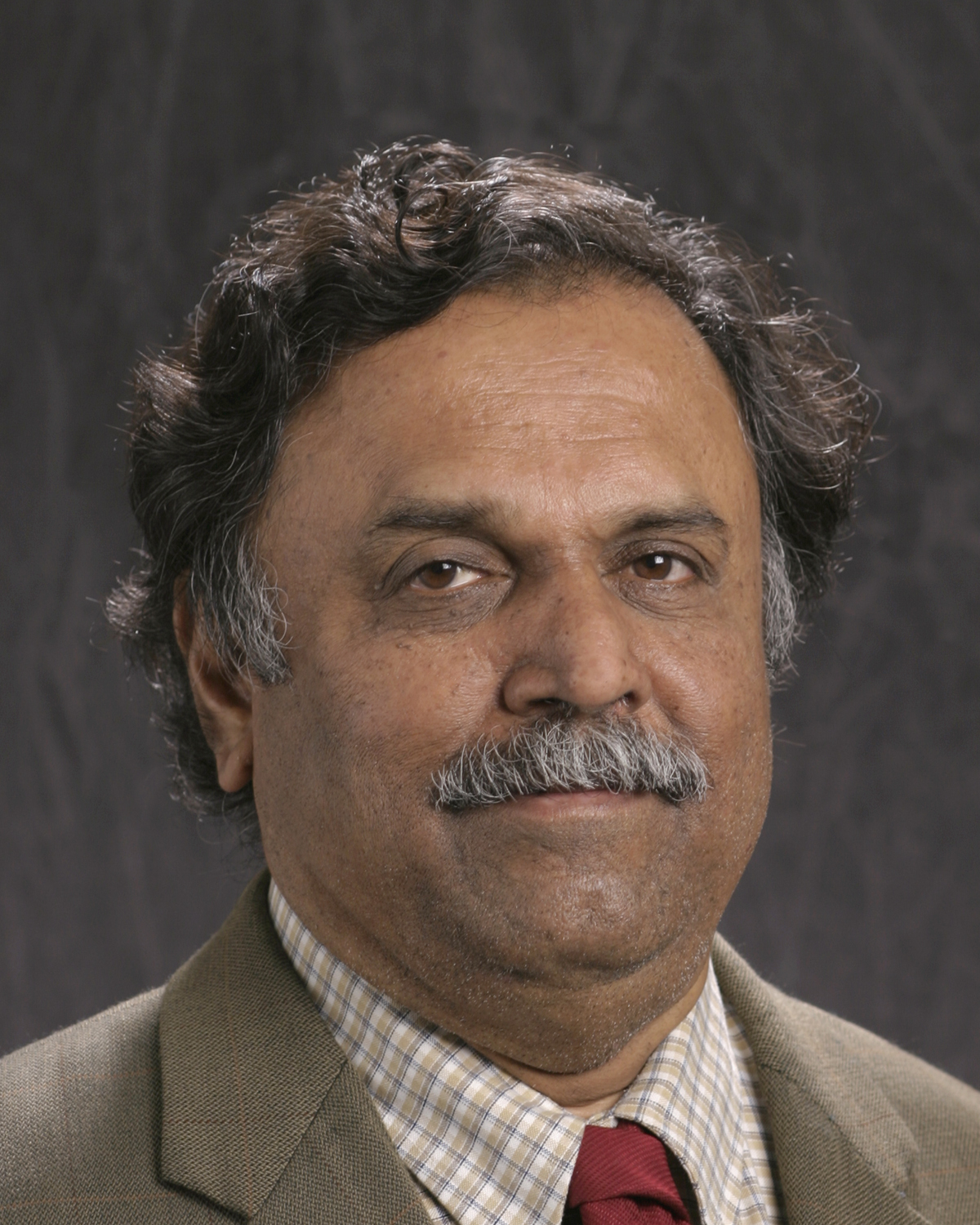}
    \vspace{-10pt}
\end{wrapfigure}
S. S. Iyengar is currently the Distinguished University Professor, Founding Director of the Discovery Lab and Director of the US Army funded Center of Excellence in Digital Forensics at FIU. He has been involved with research and education in high-performance intelligent systems, Data Science and Machine Learning Algorithms, Sensor Fusion, Data Mining, and Intelligent Systems. Since receiving his Ph.D. degree in 1974 from MSU, USA, he has directed over 65 Ph.D. students, many postdocs, and many research undergraduate students. He has published more than 600 research papers, has authored/co-authored and edited 32 books, and holds various patents. He has served on many scientific committees and panels worldwide and has served as the editor/guest editor of various IEEE and ACM journals.

His books are published by MIT Press, John Wiley and Sons, CRC Press, Prentice Hall, Springer Verlag, IEEE Computer Society Press, etc. More recently in Spring 2021, Dr. Iyengar in collaboration with HBCUs was awarded \$2.25M in funding for setting up a Digital Forensics Center of Excellence over a period of 5 years (2021--2026). He received an honorary Doctor of Science from Poznan University of Technology in Poland in May 2023. Dr. Iyengar is a Member of the European Academy of Sciences, a Life Fellow of the Institute of Electrical and Electronics Engineers (IEEE), a Fellow of the Association of Computing Machinery (ACM), a Fellow of the American Association for the Advancement of Science (AAAS), a Fellow of the Society for Design and Process Science (SDPS), and a Fellow of the American Institute for Medical and Biological Engineering (AIMBE).

\vspace{8pt}

\subsection*{Jerry Miller}
\begin{wrapfigure}{l}{1.1in}
    \vspace{-10pt}
    \includegraphics[width=1.05in]{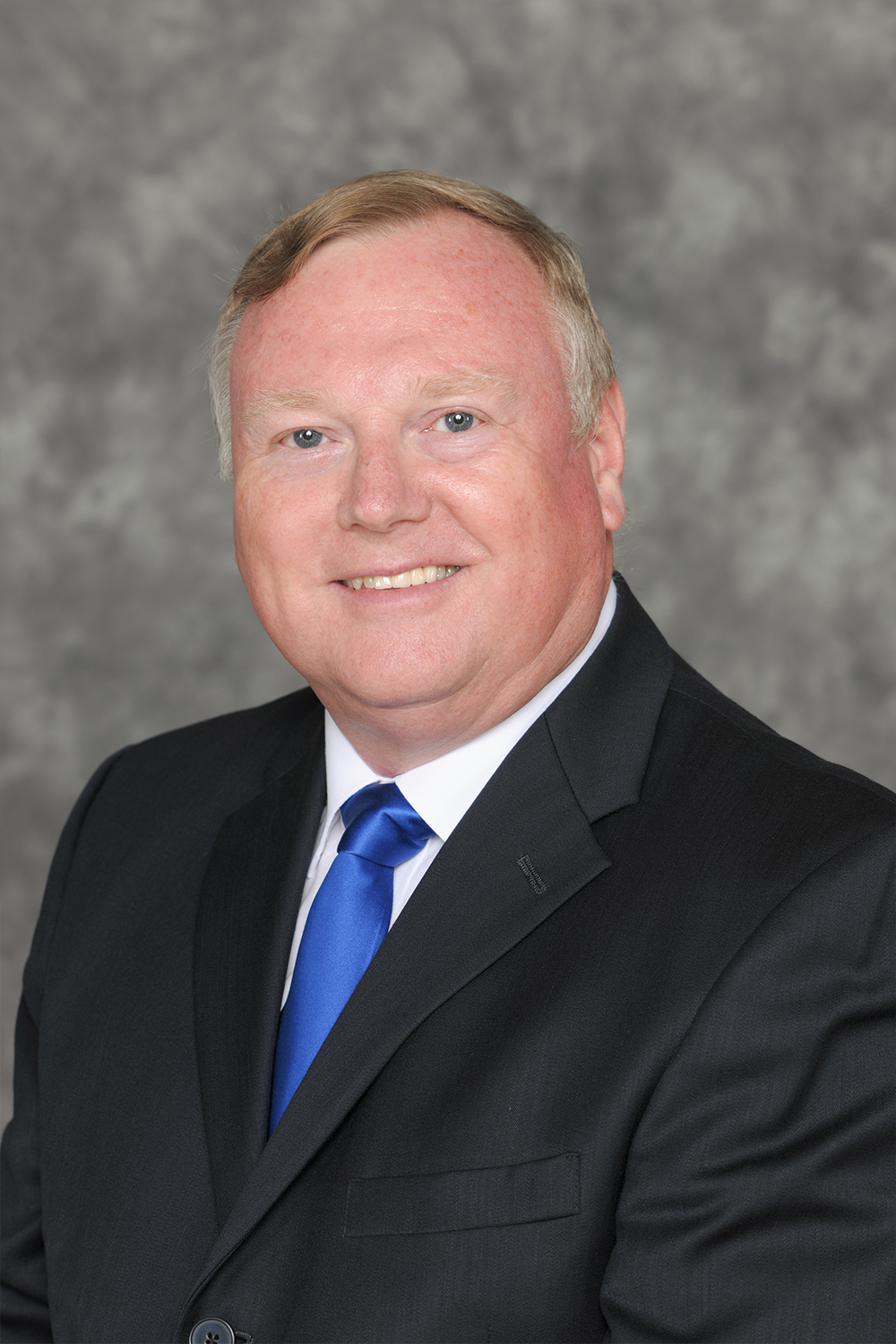}
    \vspace{-10pt}
\end{wrapfigure}
Jerry Miller is an assistant professor in the Department of Computing and Information Sciences at Florida Agricultural and Mechanical University. Prior to joining the Florida A\&M faculty, Jerry served as Associate Director in several positions at Florida International University (FIU). He joined FIU’s Applied Research Center in 2006 where he conducted research in renewable energy and global security. Later, he joined Dr. S.S. Iyengar in FIU’s Knight Foundation School of Computing and Information Sciences. During this time, he continued his education, earning his Ph.D. in computer science and conducting research in computer network security, information security, and digital forensics with Professor Iyengar.

Before entering academia, Colonel Miller served in the United States Air Force as a rescue and special operations helicopter pilot, foreign area officer, and planning/programming/budgeting officer. He is an experienced leader and aviator with an extensive background in international human rights, as well as Latin American and Pacific Region foreign policy. He has lived and worked in many countries throughout the world but always finds his way home to the majestic live oaks of Florida.

\vspace{8pt}

\subsection*{Pronab Mohanty}
\begin{wrapfigure}{l}{1.1in}
    \vspace{-10pt}
    \includegraphics[width=1.05in]{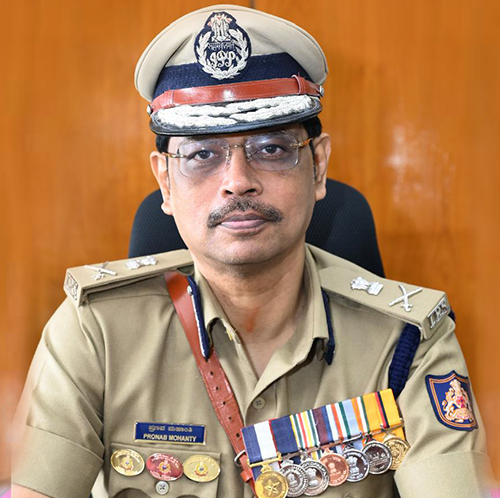}
    \vspace{-10pt}
\end{wrapfigure}
Pronab Mohanty is a highly accomplished officer from the 1994 batch of the Indian Police Service (IPS), currently serving as the Director General of Police, Cyber Economics and Narcotics (CEN). With over three decades of distinguished service, he has established himself as a leading expert in addressing complex challenges in cybercrime, economic offenses, and narcotics control.

Dr. Mohanty holds a bachelor's degree in Electronics and Instrumentation Engineering and a master's degree in Civil and Criminal Law from Utkal University, India. Furthering his academic pursuits, he earned a Ph.D. in Computer Science from the University of Minnesota, USA, where his research focused on leveraging advanced computing techniques to combat cybercrime and enhance digital security frameworks.

Throughout his career, Dr. Mohanty has demonstrated a deep commitment to public service, particularly in the realms of human trafficking, cybercrime prevention, women and children's equity, and fostering police-public partnerships. As a Deputy Inspector General of Police with the Central Bureau of Investigation (CBI) in Orissa, he spearheaded multiple groundbreaking investigations that earned national recognition for their impact and innovative approaches. A passionate advocate for modernizing law enforcement, Dr. Mohanty has led numerous initiatives to integrate technology into policing strategies. His work has not only contributed to improving cybercrime detection and prevention but has also strengthened the institutional response to crimes targeting vulnerable populations.

\vspace{8pt}

\subsection*{Naveen Kumar Chaudhary}
\begin{wrapfigure}{l}{1.1in}
    \vspace{-10pt}
    \includegraphics[width=1.05in]{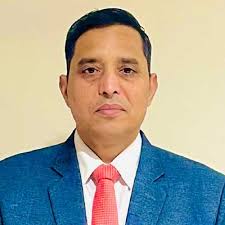}
    \vspace{-10pt}
\end{wrapfigure}
Naveen Kumar Chaudhary has been a professor of Cyber Security at the National Forensic Sciences University in Gandhinagar, Gujarat, India since 2019. He is also a Courtesy Research Professor at the Knight Foundation School of Computing and Information Sciences at Florida International University, Miami, Florida. He holds a Bachelor of Technology degree in Information Technology \& Telecommunication Engineering and Master of Engineering degree in Digital Communication. He earned his Ph.D. in Engineering and advanced certifications in Cyber and Network security. His extensive experience spans more than 25 years in engineering education, research, and government. He has steered many cutting-edge ICT projects and worked extensively on policy formulation in cybersecurity and e-governance. He is the recipient of a letter of appreciation for contribution towards the cause of literacy from Brent St. Denis, MP, Algoma Canada, in 1994. Dubai, SEWA award for his contribution to Cyber Security education in 2022. He also received COAS and CCOSC Commendation in 2009 and 2015, respectively for his innovation and distinguished service. He is an IEEE senior member and life member of IETE.

\end{document}